\documentclass[aps,twocolumn]{revtex4-2}
\usepackage{multirow}
\usepackage{float}
\usepackage[colorlinks]{hyperref}
\usepackage{color,graphicx,amsmath,amssymb,bbold}
\hypersetup{colorlinks,citecolor=blue,linkcolor=blue,urlcolor=blue}
\begin{document}
 \title{Electron spectra in double quantum wells of different shapes}
 \author{Piotr Garbaczewski, Vladimir A. Stephanovich and Grzegorz Engel}
 \affiliation{Institute of Physics, University of Opole, 45-052 Opole, Poland}
 \date{\today }
 \begin{abstract}
 We suggest a method for calculating electronic spectra in ordered and disordered semiconductor structures (superlattices) forming double quantum wells (QW). In our method, we represent the solution of Schr\"odinger equation for QW potential with the help of the solution of the corresponding diffusion equation. This is because the diffusion is the mechanism, which is primarily responsible for amorphization (disordering) of the QW structure, leading to so-called interface mixing. We show that the electron spectrum in such a structure depends on the shape of the quantum well, which, in turn, corresponds to an ordered or disordered structure. Namely, in a disordered substance, QW typically has smooth edges, while in ordered one it has an abrupt, rectangular shape. The present results are relevant for the heterostructures like GaAs/AlGaAs, GaN/AlGaN, HgCdTe/CdTe, ZnSe/ZnMnSe, Si/SiGe, etc., which may be used in high-end electronics,  flexible electronics, spintronics, optoelectronics, and energy harvesting applications. 
\end{abstract}
 \maketitle

\section{Introduction}

The use of semiconductor quantum wells (QWs) to improve the functionality of different electronic, spintronic and optoelectronic devices has been the subject of a number of studies \cite{qwd,stsn, fzds2004,bibes}. QWs are thin semiconductor layers, confining electrons or holes in one spatial dimension. Such layered structures can be fabricated to a high precision by modern epitaxial techniques. This means that one can easily control many their physical properties and quantum effects in particular. The above quantum confinement of charge carriers defines many of QWs salient experimental features \cite{qwd,stsn,bibes}. A typical quantum well consists of thin layers (around 50 atomic layers thick) of one semiconductor ("well" material) sandwiched between other semiconductor or "barrier" material. One of the examples of a "well" material is GaAs, while the "barrier" material is usually a semiconductor with a larger band gap like AlGaAs, see e.g. \cite{qwd}. As a result of quantum confinement in such a quantum well, the discrete energy levels (so-called sub-bands) occur, which substantially change the physical (like optical) properties of such heterostructures as compared to the pure bulk "well" or "barrier" material. 

From a fundamental point of view, QW heterostructures can be regarded as an artificial "playground", where one can explore various quantum mechanical effects, which are inaccessible for free atoms confined in the traps \cite{qwd}. For example, the above QWs can confine charge carriers in very small (compared to atomic gases) distances which permits to apply effectively high electric and magnetic fields to them. Latter fields can be made that high due to the thinness of semiconductor layers in QW structures. On the other hand, many physical effects of QW heterostructures occur at room temperature and thus can be used in real devices.

When QWs are brought close to each other, their coupling occurs through the charge carrier tunnelling, which can take place in two- or multi-well structures. The novel physical (primarily optical) properties of such multilayer structures, appear due to various electronic transitions between above subbands, which comprise discrete levels in the case of one, two and small numbers of coupled QWs. One of the most important applications of the above multi-well structures is so-called quantum cascade lasers \cite{faist}, where the lasing effect is enhanced greatly due to the resonant tunnelling of the electron between adjacent QWs, accompanied by the cascaded intersubband transitions. In order to predict correctly the behaviour of above devices, one needs to know the detailed energy spectrum of real quantum multi-well structures. 
 
The calculations of electron energy spectra in QW structures are usually performed in the approximation where QW edges (which are actually the conduction and valence band offsets, see e.g. \cite{qwd}) are taken as step functions. At the same time, the different kinds of disorder influence the shape of quantum wells. The origin of such disorder is local stochastic variation of chemical composition during heterostructure fabrication, leading to the deviation of wells boundaries from ideal step functions. The random diffusion of impurities at the interfaces is also a strong possibility, especially at elevated temperatures used in the fabrication process. Such diffusion will "smear" the QW profile, leading sometimes to drastic changes of the electronic spectrum as compared to that in the perfectly abrupt case. This, in turn, will affect the device characteristics like slope efficiency in lasers, energy conversion efficiency in the solar cells and the overall operation lifetime of any device. As the above device characteristics strongly depend on the QW materials purity, the quest to obtain pure multi QW structures is ongoing now \cite{te1,te2,te3,te4}. That is why the major efforts are concentrated on the fabrication of high-quality QW structures on the base, most frequently of the alloy Ga$_{1-x}$Al$_x$As \cite{te3,te4}. To fabricate the latter structures with predefined physical properties, the combination of technological "trial and error" approaches along with theoretical ones, based on different density functional theory (DFT) varieties, are utilized \cite{dft1,dftpccp}. The main feature of all DFT varieties is very frequently the requirement of sizable experimental input so that they are in fact based on phenomenological approximations of many-body potentials in so-called supercells, see, e.g., Ref. \cite{dftpccp}. The second important feature of such approaches is that they are highly (sometimes prohibitively) computer-intensive. A special place here belongs to the effects of the disorder, which is almost impossible (except a couple of empirical recipes \cite{dft1}) to account for in the DFT approaches. 

In the present paper we consider the energy spectrum of a charge carrier in a double QW structure with respect to the effects of disorder.  Our method is to derive the solutions of the Schr\"odinger equation from those of Fokker-Planck (diffusion in an external potential) equation \cite{pavl}, which describes the probability density function (pdf) in the above random diffusion process in the QW vicinity. The physical reason for that is that the diffusion is responsible primarily for amorphization (disordering) of the QW structure, leading to the smoothing of QW profiles. In that sense our method considers the disorder self-consistently and may be regarded as "the method of self-consistent disorder". The transformation of the initial Fokker-Planck equation is accomplished by the Schr\"odinger semigroup exp$(-t{\cal H}/\hbar)$, where $\hbar$ is Planck constant, $t$ is time and ${\cal H}$ is a system Hamiltonian \cite{pavl,risken,jph1}. First, this transformation facilitates the solution of the Schr\"odinger equation in the smooth, disorder-influenced QW potential, which is a nontrivial task. Second, it will permit us to trace the influence of disorder-related QW profile smearing on the energy spectrum of the charge carriers.  We discuss also the experimental implications of the above smearing. Our method can be extended for the consideration of the excitonic effects, which can significantly influence the optical absorption coefficient and other properties, especially at room temperatures.

\section{Methodology}

\subsection{General formalism}

The easiest way to understand the basic properties of an electron (or hole) in a quantum well structure is to consider corresponding stationary Schr\"odinger equation in one dimension 
\begin{equation} \label{mej1}
	-\frac{\hbar^2}{2m^*}\frac{d^2\psi}{dx^2}+V(x)\psi=E\psi,
\end{equation}
where $m^*$ is an electron (hole) effective mass, $V(x)$ is the (for instance single or double-well) potential, seen by the electron along the $x$ direction and $E$ and $\psi$ are the eigenenergy and eigenfunction, giving the solution to Schr\"odinger equation \eqref{mej1}. The eigenenergies $E$ can further be used to calculate the observable characteristics of quantum well structures like optical absorption coefficient, which is proportional to the density of states.

As a QW structure is in fact "embedded" in a 3D sample and in view of the fact that the above stochastic processes at the interfaces are three-dimensional and time-dependent, we begin with 3D time-dependent Schr\"odinger equation
 \begin{equation} \label{mej1a}
	-i\hbar \partial _t \Psi ({\bf r},t)= {\cal H} \Psi({\bf r},t), 
\end{equation}
where $\partial _t$ stands for time derivative and ${\cal H}$ is the Hamiltonian
 \begin{equation} \label{mej1b}
 {\cal H}=-\frac{\hbar^2}{2m^*}\Delta+V({\bf r}),
 \end{equation}
 where $\Delta$ is a Laplacian and $V({\bf r})$ is now a potential taking into account the deviation (due to diffusion, for instance) of the QW shape from 1D case. Note that the standard substitution $\Psi({\bf r},t)=\psi({\bf r})e^{iEt/\hbar}$ generates equation \eqref{mej1} in 1D case.
 
To relate the eigenfunction of the Schr\"odinger equation \eqref{mej1a} to the pdf $\rho({\bf r},t)$ of the "stochastic diffusion" in a   heterostructure, we consider the Fokker-Planck (Smoluchowski) equation, which has the form \cite{risken}
\begin{equation}\label{mej2}
\partial _t \rho  = D \Delta \rho - \nabla [b({\bf r}) \rho],
\end{equation}
where $D=\tau k_BT/m^*$ is diffusion coefficient. The latter coefficient is considered to be independent of time and coordinates. Here, the usual Einstein relation for $D$ has been used \cite{risken} with $m^*$ being the above effective mass, $\tau$ is a mean relaxation time of a system, $T$ is a temperature and $k_B$ is Boltzmann constant. Also, $b({\bf r})= - (\tau/m^*)\nabla U({\bf r})$ is a force field, related to (confining) potential $U({\bf r})$, in which the diffusion (or any of the above random processes, described by a stochastic Langevin equation, see \cite{risken,lang} and references therein) occurs. As in real structures, the diffusion occurs in three spatial dimensions, the above form \eqref{mej2} reflects this fact. If necessary, the form \eqref{mej2} can easily be recast to the lower dimensions.

It is well-known that the Fokker-Planck equation \eqref{mej2} takes the initial $\rho({\bf r},t=0)\equiv \rho_0({\bf r})$ to an asymptotic $t \to \infty$ stationary pdf of the Boltzmann form 
\begin{eqnarray}
\rho _*({\bf r})= \frac{1}{Z}\exp\left(- \frac{U({\bf r})}{k_BT}\right), \label{mej3}\\  
Z= \int \exp\left(- \frac{U({\bf r})}{k_BT}\right)d^3r, \nonumber
\end{eqnarray}
where $Z$ is the normalization constant. 

The "bridge" between the initial Schr\"odinger equation \eqref{mej1a} and the Fokker-Planck one \eqref{mej2} can be set if we note that the eigenvalues of the Hamiltonian \eqref{mej1a}, up to an inverted sign, are similar to those of the operators ${\cal L}^*$ and ${\cal L}$, where ${\cal L}^*\rho=D \Delta \rho - \nabla [b({\bf r}) \rho]$ is the Fokker-Planck operator in \eqref{mej2}, while the adjoint operator ${\cal L}$ is the diffusion generator of the corresponding stochastic process \cite{pavl}. To be specific, having a stationary pdf $\rho _*$ \eqref{mej3}, one can  transform the Smoluchowski-Fokker-Planck evolution  $\exp(t{\cal L}^*/\hbar)$, to the Schr\"{o}dinger semigroup  $\exp(-t {\cal H}/\hbar)$, see e.g.   \cite{jph,jph1}. A classic factorization  \cite{risken} of $\rho ({\bf r},t)= \Psi ({\bf r},it) \rho _*^{1/2}({\bf r})$  ($\Psi ({\bf r},it)$ is the solution of \eqref{mej1a} with respect to $t \to it$) allows to map  the Fokker-Planck  dynamics into the generalized  Schr\"{o}dinger-type problem, which reads   
   \begin{equation} \label{mej4}
   -\hbar \partial _t \Psi = {\cal H} \Psi.
   \end{equation}
The above shows that the Fokker-Planck equation reduces to Schr\"odinger one for imaginary times, which imply that instead 
of oscillatory behavior of the wave function $\sim e^{iEt/\hbar}$, here we have true exponential relaxation. This means that now the stationary (at $t \to \infty$) pdf \eqref{mej3} is actually achievable. For that condition to fulfill \cite{jph, faris}, the potential $V({\bf r})$ in the Hamiltonian \eqref{mej1b} should be related to $U({\bf r})$ from Eqs. \eqref{mej2},\eqref{mej3} 
\begin{equation} \label{mej5}
V({\bf r}) =  \hbar D{\frac{\Delta \rho _*^{1/2}}{\rho _*^{1/2}}} = \frac{\hbar D}{2k_BT} \left(\frac{[\nabla U({\bf r})]^2}{2k_BT} -\Delta U({\bf r})\right).
\end{equation}
In other words, the   relaxation process $\rho ({\bf r},t) \to \rho_*({\bf r})$ is paralleled by the relaxation $\Psi _0({\bf r}) \to \Psi ({\bf r},it) \to \rho _*^{1/2}({\bf r})$. It is seen from \eqref{mej4} that ${\cal H} \rho_*^{1/2}= 0$ so that $\rho _*^{1/2}$ is a legitimate {\it zero energy} eigenstate of ${\cal H}$. Below we shall see that the knowledge of latter discrete eigenstate of the Hamiltonian \eqref{mej1b} (with respect to \eqref{mej5}) will permit us to classify the eigenstates in the "disordered" (i.e. smooth) wells and by this virtue to trace the influence of disorder (QW boundaries smoothness) on the electron energy spectrum in such structure. As the Fokker-Planck equation describes the diffusion, which through the expression \eqref{mej5} influences the QW potential in the Schr\"odinger equation \eqref{mej1b}, our treatment of the diffusion-generated disorder is in fact self-consistent. 

As we are primarily interested here in the properties of (effectively one-dimensional) potential wells in our structure, we consider following {\em{ansats}} for the stationary wave function $\Psi$ in \eqref{mej4}
\begin{equation}\label{zux1}
\Psi (\mathbf k)\equiv \Psi_n({\mathbf k})=\psi_n(x)u({\mathbf r}_\perp)e^{i{\mathbf k}{\mathbf r}_\perp},
\end{equation}
where $n$ is energy level index, which enumerates the discrete states in a QW, ${\mathbf k}$ and ${\mathbf r}_\perp$ are the wave vector and coordinate in the $yz$ plane, perpendicular to QW growth direction and $u({\mathbf r}_\perp)$ is the Bloch function. This is because the Hamiltonian \eqref{mej4} is translationally invariant in the $yz$ plane. As here we consider only a discrete spectrum in a well, below we omit the level index $n$ in the "well" wave function $\psi_n(x)$,
restoring it only where it is necessary.  

One more interesting observation is in place here. Namely, if the potential $U({\bf r})$ for the diffusing charge carrier is monomial (i.e. single well), the induced potential $V({\bf r})$ in the Schr\"odinger equation is usually smooth double-well. 
This demonstrates also the essence of the suggested method, where the double - well potential in the Schr\"odinger equation \eqref{mej1b} was not introduced initially, but had rather been derived from the Fokker-Planck equation, which describes the diffusion process in the confining potential, which is usually the case for technological processes of QW structures fabrication.  

As this effect can best be seen in the 1D case, we take the family of potentials $U$ in the form, using commonly to parametrize the monomial, single-well potentials  

\begin{equation} \label{mej6}
U(x)= {\frac{\kappa_m}m} x^m,\ \kappa_m=m,\ m^2.  
\end{equation}
Here, the coefficient $\kappa_m$ (which depends on integer parameter $m$) parameterizes the above family of potentials \eqref{mej6} in a unified way. For instance, the specific form $U(x)=x^{10}$ corresponds to $m=10$ and $\kappa_m=m$. On the other hand, $U(x)=10x^{10}$ corresponds to the same $m=10$ and $\kappa_m=m^2$.
 
In the expression \eqref{mej6}, we suppress all dimensional parameters, which can be easily restored in numerical estimations for the actual semiconductors. The application of the formula \eqref{mej5} (in 1D case and adopted dimensionless variables) to the $U(x)$ \eqref{mej6} immediately generates the following expression for $V(x)$ 

\begin{equation}\label{mej7}
 V(x) = ax^{2m-2} - bx^{m-2},
\end{equation}
where  $a= \kappa _m^2/4$ and   $b= \kappa_m  (m-1)/2$. The comparison of the expressions \eqref{mej6} and \eqref{mej7} shows that the case $m=2$ is distinct. Really, the $U(x) \sim x^2$ in \eqref{mej6} implies (up to a constant) $U(x) \sim x^2$ also. This is related to the fact that the quadratic potential in quantum mechanics is special as it has accidental degeneracy (along with Coulomb potential in the hydrogenic problem), which leads to the above result, see, e.g. \cite{land3}. This effect is important only in the consideration of the excitonic effects, where the case $m=2$ can be special. As we postpone the consideration of the excitonic effects for the future work, below we shall consider only $m>2$ cases.

\begin{figure}[h]
\begin{center}
\includegraphics [width=0.97\columnwidth] {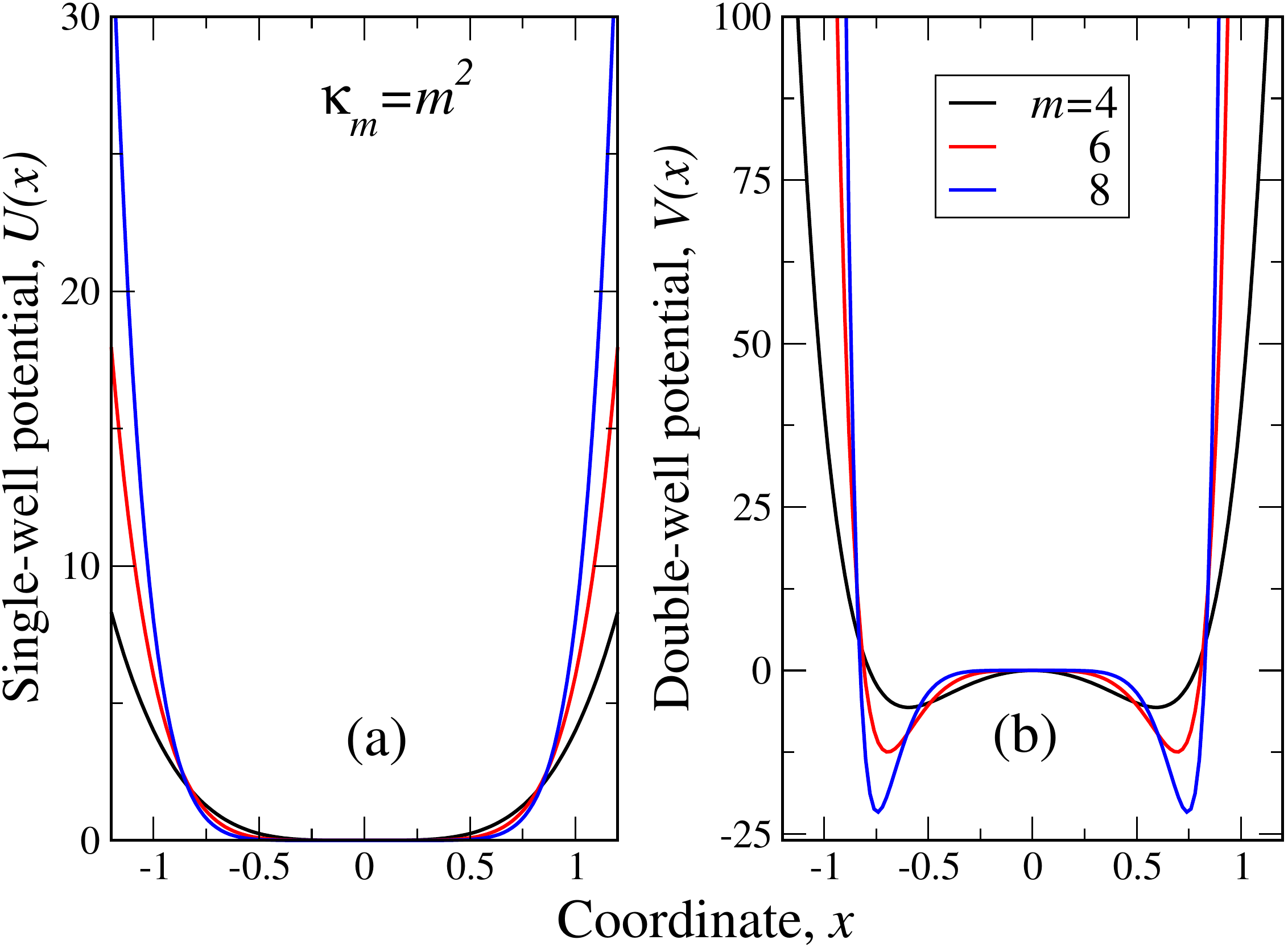}
\caption{Single-well \eqref{mej6} (panel (a)) and corresponding double-well \eqref{mej7} (panel (b)) potentials for $\kappa_m=m^2$ and different $m$'s, shown in the legend. The behavior of the potentials for $\kappa_m=m$ is qualitatively similar.}
\label{fig:gi1}
\end{center}
\end{figure}

The potential $V(x)$ \eqref{mej7} for $\kappa_m=m^2$ and several $m>2$ along with the corresponding monomial potentials $U(x)$ \eqref{mej6} are reported in Fig. \ref{fig:gi1}. The shape of the above potentials for $\kappa_m=m$ is qualitatively similar. It is seen that as $m$ increases, the monomial potential $U(x)$ becomes progressively steeper so that at $m \to \infty$ it tends to the square infinite potential well. This shows that the diffusion process in the very steep potentials does not generate essential disorder as such highly confining potential "channels" the particle flux, making it almost deterministic. At the same time, as it is seen from the Fig.\ref{fig:gi1}(b), in the corresponding double-well potential at higher $m$'s the minima become deeper and narrower, while the "tails" at $|x|>>1$ become also steeper. In other words, at large $m$ the double-well potential $V(x)$ would also resemble the infinite rectangular single well. This confirms the above statement, that very steep potential does not generate the essential disorder. The only exception is that at $m \to \infty$ the curve $V(x)$ will have the "needle-like", infinitely deep minima around $|x|=1$ or in dimensional units $x=\pm L$, where $L$ is the technological width of the "ordered" quantum well. To have better insights in the "order-disorder" transition in our model, below we shall use both above limiting $m \to \infty$ procedure and the approximation of smooth double-well potential \eqref{mej7} by rectangular one, see \cite{bas}.   

%*****************************************************************************************
%*****************************************************************************************
The ground state function of the Hamiltonian \eqref{mej1b} $\rho_*^{1/2}(x)= (1/\sqrt{Z})  \exp [U(x)/2]$ is unimodal (i.e. does not have nodes as it should be for ground state function \cite{land3,bas}) with a maximum at $x=0$. The knowledge of this ground state function facilitates tremendously the analysis of the spectra of both smooth (disordered) and rectangular (ordered) potentials. The spectra of such potentials had been analyzed approximately in Refs. \cite{bas,baner}. Note that in spite that we approximate the "ragged" QW profiles in disordered heterostructures by smooth ones, our numerical simulations show that this does not change any of the qualitative conclusions.  

Having the above information in our disposal, now we can formulate our algorithm. Specifically, as the actual shape of the quantum wells in a semiconductor structure depends strongly on the history of its fabrication, or equivalently, of the details of the 
stochastic migration of the chemical species, we begin with the consideration of latter process in the framework of the equation \eqref{mej2}. Having its pdf $\rho_*(x)$, we can deduce the ground state eigenfunction of the QW Hamiltonian \eqref{mej1b} and restore numerically its whole spectrum (see \cite{turbiner}) as well as the shape of a quantum well from the equation \eqref{mej5}. The information about the above spectrum can be used to calculate the physical properties like optical absorption coefficients as well as to construct the transition probability densities of the  pertinent diffusion process, see, e.g., \cite{risken,pavl}.
 
\subsection{The approximation of the smooth double-well potential by a rectangular one. Comparison of the spectra.}

\begin{figure}[h]
\begin{center}
\centering
\includegraphics[width=0.99\columnwidth]{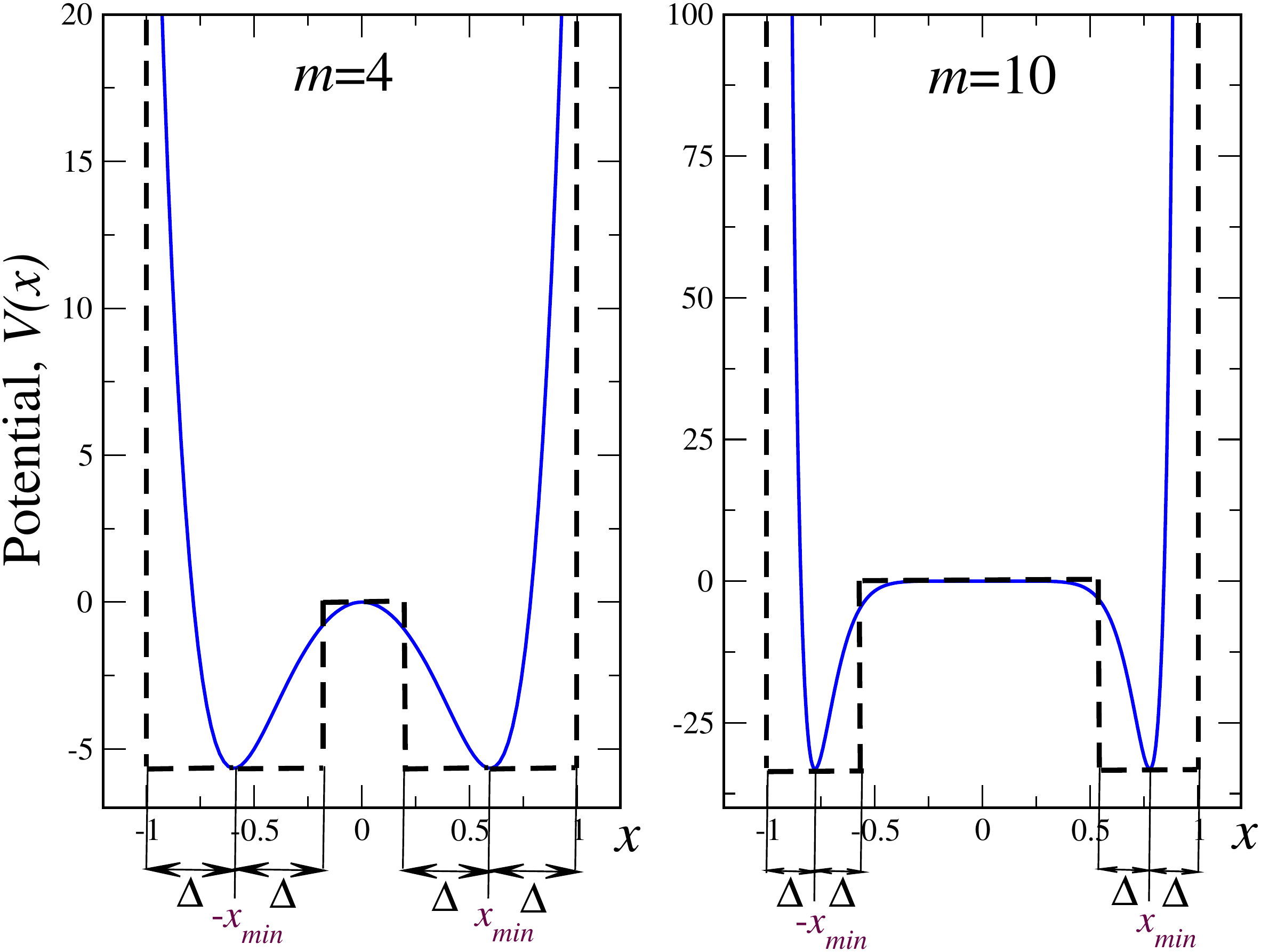}
\caption{Exemplary fitting of $m=4$ (left panel) and $m=10$ (right panel) potentials \eqref{mej7} for $\kappa_m=m^2$ to  suitable rectangular double wells.  Here, $\Delta$ stands for a distance between the nearby endpoint $\pm 1$ and abscissa of the potential minimum $x_{min}$, i.e. $\Delta=|1-|x_{min}||$. In this case, the rectangular barrier ("plateau" in the vicinity of $x=0$) width is $2(1-2\Delta)$.} \label{fig:gi2}
\end{center}
\end{figure}

To understand the disorder influence on the QW spectra, it is instructive to approximate the smooth double-well potential \eqref{mej7} by the best fit rectangular one (Fig. \ref{fig:gi2}) and compare the spectra in both cases. This approximation procedure can be performed along the lines of Ref. \cite{bas}. It needs the knowledge of the behavior of the minima of the potentials \eqref{mej7} as $m \to \infty$. 

The minima of the potential \eqref{mej7} can be found in the usual way. We have   

\begin{eqnarray}
x_{min}= \pm \left[ {\frac{ m-2}{\kappa _m}} \right]^{1/m}, \nonumber \\
V(x_{min})=-\frac 14 m(m-2) x_{min}^{-2}.  \label{mej8}
\end{eqnarray}

%*************************************************************************************
%*************************************************************************************

For large $m$, the formula  \eqref{mej8} can be rewritten as follows:

\begin{eqnarray}
&&|x_{min}| =  \exp \left \{ {\frac{1}m}  \left[ \ln {\frac{m}{\kappa_m }}  + \ln \left(1 - {\frac{2}m}\right)\right] \right \} _{x \to \infty}\approx \nonumber \\ \nonumber \\ 
&&\left\{\begin{array}{ll}
1 - \frac{\ln m}{m}+\frac{1}{m^2}\left(\frac 12\ln^2m-2\right)+...,& \kappa_m=m^2 \\ \\
1-\frac{2}{m^2}+...,& \kappa_m=m.	
\end{array}
\right. \label{mej9}
\end{eqnarray}
The asymptotics \eqref{mej9} shows immediately that at $m \to \infty$ $x_{min} \to \pm 1$. It also follows from \eqref{mej9} that at $m \to \infty$ $x_{min}^{-2} \to 1$ so that in this limiting case $V(x_{min}) \to -\infty$, see expression \eqref{mej8}. This behavior is in accord with Fig. \ref{fig:gi1} and its discussion.

The idea behind the approximation of the smooth potential with square one is that the ground state $\rho_*^{1/2}$ (where $\rho_*$ is given by Eq. \eqref{mej3}) of the Hamiltonian \eqref{mej1b} with the smooth double-well potential \eqref{mej7} should be as close as possible to that of a square approximant. For that we adopt the method, used in Ref. \cite{bas} for the ammonia molecule, where the initial smooth double-well potential has been approximated by a sequence  of  rectangular  double well systems, with adjustable barrier heights and widths. Note that in our case, the rectangular potentials are the best approximants for those with $\kappa_m=m^2$, the approximation for $\kappa_m=m$ requires more than one "double-well rectangle".  
The only difference is the above vertical shift \cite{bas,blinder,lopez,jelic} of the potential so that the "plateau" between wells is now coincides with the maximum of the initial potential, which, in turn, occurs at $x=0$, see Fig. \ref{fig:gi2}. Such procedure also gives the coincidence of the minima (well depths) of the initial smooth potential and its square approximant. The explicit outcome of such approximation procedure is the possibility to obtain analytically (i.e. without explicit numerical solution of the Schr\"odinger equation) the whole spectrum of the Hamiltonian \eqref{mej1b} with such square double-well potential, see e.g. \cite{blinder}. 

\subsection{The spectra for rectangular and smooth cases}

As we have seen above, the Hamiltonian \eqref{mej1b} with the smooth double-well potential \eqref{mej7} admits the exact solution $\rho_*^{1/2}$ \eqref{mej3}, corresponding to zero eigenvalue. To be specific, for the dimensionless 1D version of the Hamiltonian \eqref{mej1b} 
\begin{equation} \label{mej10}
{\mathcal H}_0=-\frac{d^2}{dx^2}+V(x),
\end{equation} 
($V(x)$ is determined by \eqref{mej7}), the solution, corresponding to eigenvalue zero ($n=0$), has the form
\begin{equation}\label{mej11}
\psi_0(x)\equiv \sqrt{\rho_*(x)}=Ae^{-\frac{U(x)}{2}},
\end{equation}
where $U(x)$ is defined by \eqref{mej6} and $A$ is the normalization constant. Explicitly
\begin{equation}\label{dr1}
\psi_0(x)=\sqrt{\frac{m}{2\Gamma\left(\frac 1m\right)}}\left(\frac{\kappa_m}{m}\right)^{\frac{1}{2m}}e^{-\frac{\kappa_m}{2m}x^m},
\end{equation}
where $\Gamma(x)$ is gamma-function \cite{abr}. Our numerical simulations for the potential \eqref{mej7} show that such solution corresponds to the lowest eigenvalue, i.e. to the ground state. This is not the case for its rectangular approximants (perfectly ordered case, see Fig. \ref{fig:gi2}), where the spectrum begins from higher value. This shows the distinctive feature of our model of "self-consistent disorder". Namely, if initial "ordered QW" has strictly positive spectrum (see Fig. \ref{fig:gi2}), where zero corresponds to the top of the barrier), after the diffusion "smears" its profile, the resulting potential starts to have zeroth eigenvalue. This obviously is reflected in the density of states and such characteristics as optical absorption coefficient $\alpha$. As this coefficient is proportional to the optical density of states $\alpha \propto \sqrt{E-E_g}$ ($E_g$ is a band gap width of a semiconductor), the above fact has clear experimental and technological implication. Namely, the "disordering" of a QW structure by the diffusion leads to the lowering of the absorption edge, which in this case becomes exactly equal to $E_g$.  
Below we are going to discuss this in more details. We note here, that in our consideration all the energies are  referred to $E_g$. 

In the ordered case of rectangular double-well potentials, the spectrum of corresponding Hamiltonian can be obtained analytically except for the numerical solution of transcendental equations for continuity conditions at the points $\pm (1-2\Delta)$. Namely, the solutions obey usual boundary conditions $\psi(\pm 1)=0$ and can be classified as symmetric (corresponding to even principal quantum numbers $n$) and antisymmetric (even $n$'s) \cite{bas}. We look for solutions in the form
\begin{subequations}
\begin{equation} \label{dr2a}
\psi(x)=\left\{
\begin{array}{c}
A\sin k(1+x), \ \ -1<x<-1+2\Delta \\
B\cosh \kappa x, \quad -1+2\Delta< x < 1-2\Delta   \\
A\sin k(1-x), \quad \quad \ 1-2\Delta <x<1.   
\end{array}
\right.
\end{equation}
(even states) and
\begin{equation} \label{dr2b}
\psi(x)=\left\{
\begin{array}{c}
-A\sin k(1+x), \ \ -1<x<-1+2\Delta \\
B\sinh \kappa x, \quad -1+2\Delta< x < 1-2\Delta   \\
A\sin k(1-x), \quad \quad \ 1-2\Delta <x<1.   
\end{array}
\right.
\end{equation}
\end{subequations}
- odd states. Here $k=\sqrt{E}$ (in dimensionless units, corresponding to $k=\sqrt{2m^*}/\hbar$ in dimensional units) and $\kappa=\sqrt{|V(x_{min})|-E}$ (also in dimensionless units). Latter definition implies that $\kappa=\sqrt{|V(x_{min})|-k^2}$.

The conditions of continuity of wave functions $\psi(x)$ at the points $\pm (1-2\Delta)$ generate following relations between constants $A$ and $B$
\begin{subequations}
\begin{equation} \label{dr3a}
B=A\frac{\sin 2k\Delta}{\cosh \kappa(1-2\delta)}
\end{equation}
(even states) and
\begin{equation} \label{dr3b}
B=A\frac{\sin 2k\Delta}{\sinh \kappa(1-2\delta)}
\end{equation}
\end{subequations}
(odd states). The relations \eqref{dr3a} and \eqref{dr3b} permit to find constant $A$ from the normalization condition 
\begin{equation} \label{dr4}
\int_{-1}^1\psi^2(x)dx=1.
\end{equation}

%*************************************************************************************
\begin{figure}[h]
\begin{center}
\centering
\includegraphics[width=0.9\columnwidth]{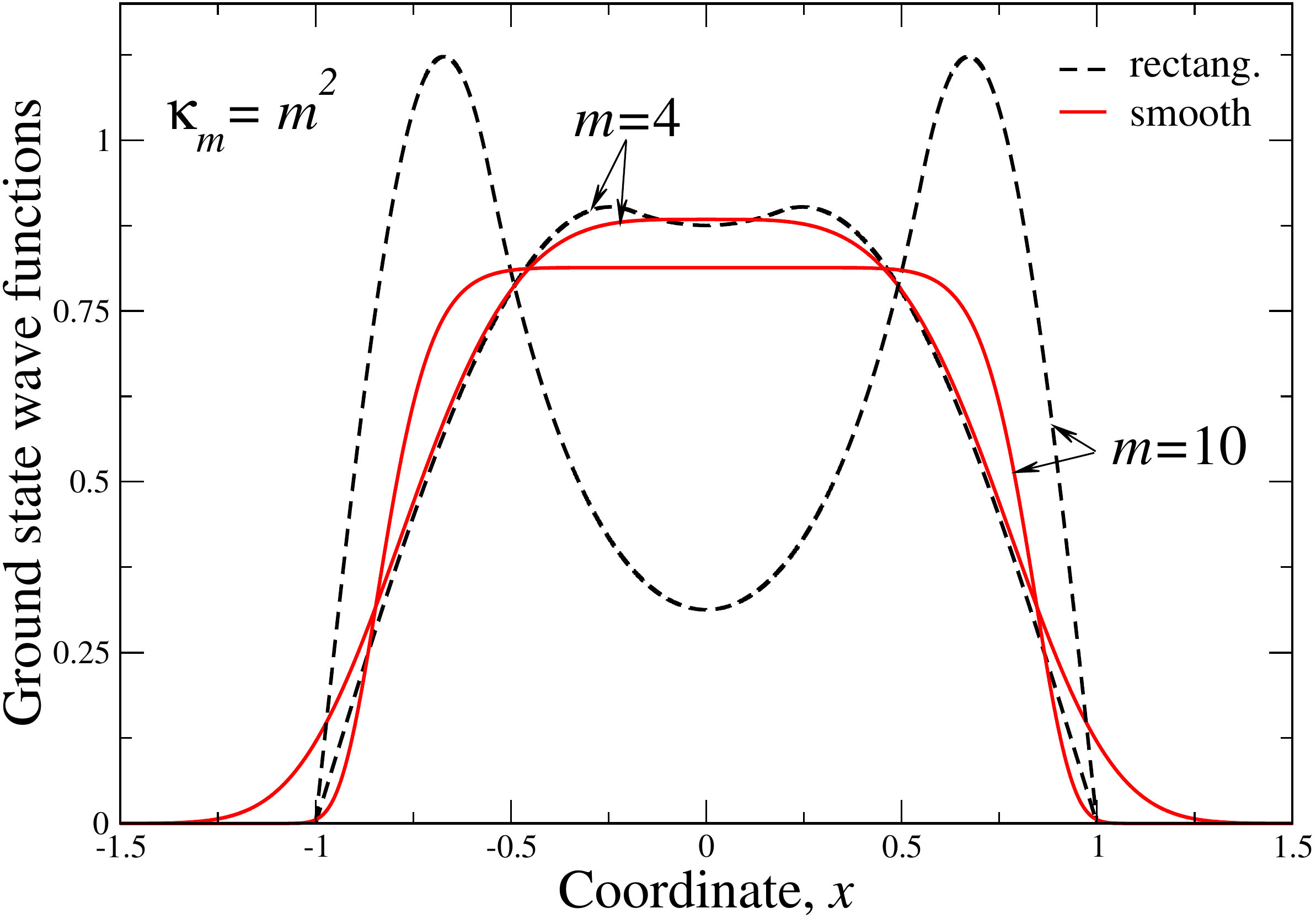}
\caption{Ground state wave functions for $m=4$ and $m=10$ for rectangular (dashed lines) and smooth (full red lines) double-well potentials, shown in Fig.\ref{fig:gi2}. Here $\kappa_m=m^2$.}  \label{fig:gi3}
\end{center}
\end{figure}
%*************************************************************************************
The usual conditions for the continuity of logarithmic derivatives $\psi'(x)/\psi(x)$ at the points $\pm (1-2\Delta)$ generate following transcendental equations for the spectrum of the particle (electron or hole) confined in our rectangular double-well potential
\begin{subequations}
\begin{equation} \label{dr5a}
k\cot 2k\Delta =-\kappa \tanh \kappa (1-2\Delta)
\end{equation}
(even states) and
\begin{equation} \label{dr5b}
k\cot 2k\Delta =-\kappa \coth \kappa (1-2\Delta)
\end{equation}
\end{subequations}
(odd states). For $m=10$ and $\kappa_m=100$ we have from \eqref{mej8} $x_{min} \approx \pm 0.7768$ and $\Delta \approx 0.2232$. This implies the wells depth $V(x_{min})\approx -33.1445$, see also right panel of Fig. \ref{fig:gi2}. Substitution of these data into \eqref{dr5a} yields ground state energy $E_0 \approx -10.26941$ and second excited state one $E_2 \approx 9.23526$. The equation \eqref{dr5b} gives the first excited state energy $E_1 \approx -9.559704$, which is, along with $E_0$, is less than zero, i.e. lies inside the wells. This means that in this case the wells in the double-well potential, are almost decoupled. The same is true for rectangular potential, shown in the left panel of Fig. \ref{fig:gi2}. In this case we have $m=4$, 
$\kappa_m=16$ so that $x_{min} \approx \pm 0.594604$, $\Delta \approx 0.405396$ and $V(x_{min})\approx -5.65685$. These parameters yield $E_0 \approx -1.32108<0$, $E_1 \approx 4.435472$ and $E_2 \approx 18.31606$. It is seen that for $m=4$ the ground state energy is closer to zero than that for $m=10$. This means that for high $m$ and $\kappa_m=m^2$ the wells in ordered, rectangular structure become progressively more decoupled. At the same time, for the corresponding smooth potentials ground state energy $E_{0s}$ is always zero (see above), which means that the disorder effectively couples quantum wells in a structure. This leads, for instance, to the lowering of the absorption edge, which can be manifested both in experiment and in optoelectronic device applications. The numerical solution of the Schr\"odinger equation with the "self-consistent potentials" from Fig. \ref{fig:gi2} yields $E_1\approx 5.474368$ and $E_2 \approx 17.814829$ for $m=4$ ($\kappa_m=16$) and $E_1\approx 4.280768$ and $E_2 \approx 16.694878$ for $m=10$ ($\kappa_m=100$). 

%*************************************************************************************
\begin{figure}[h]
\begin{center}
\centering
\includegraphics[width=0.99\columnwidth]{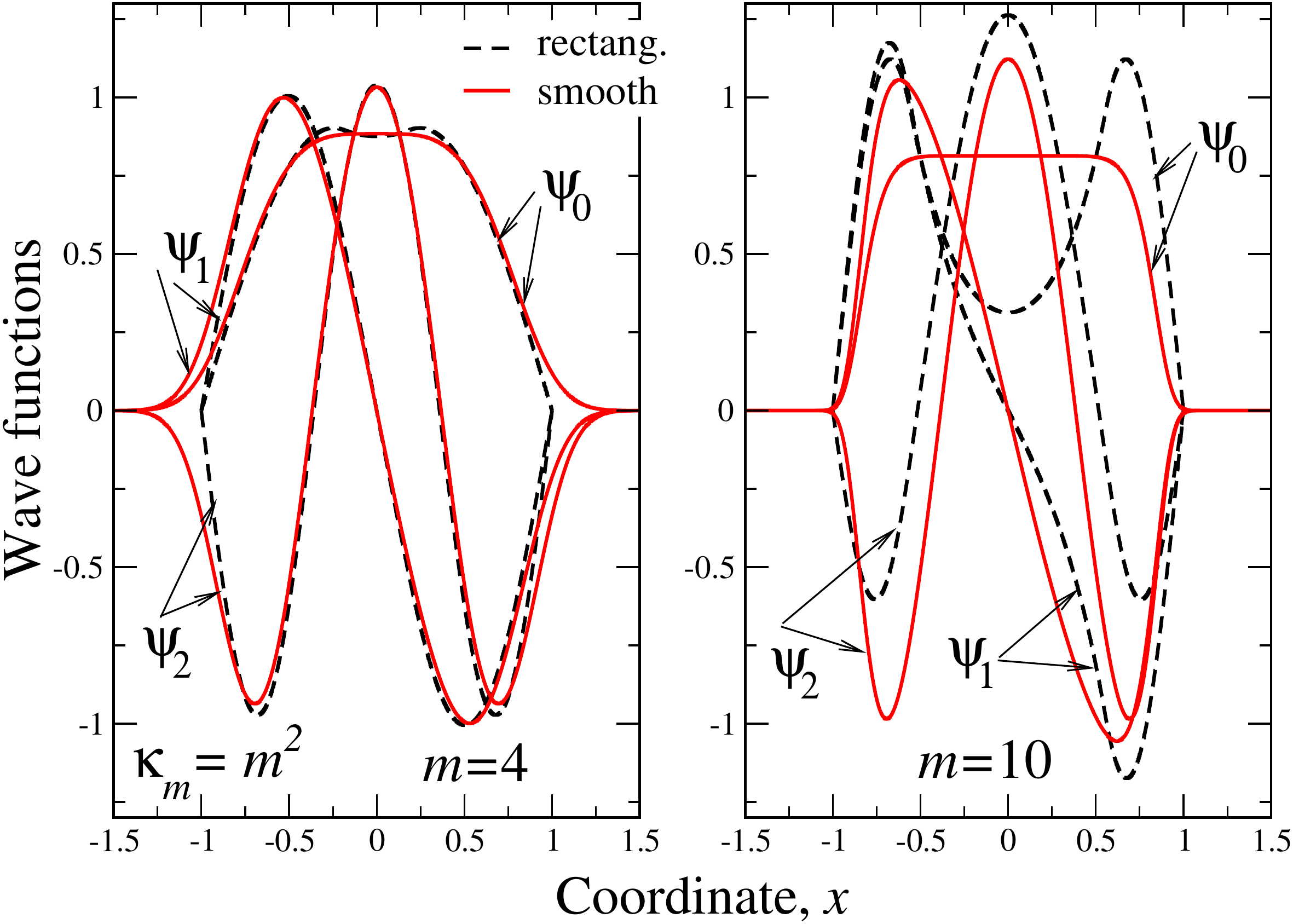}
\caption{Same as in Fig. \ref{fig:gi3}, but for ground $\psi_0$ and two lowest excited states $\psi_{1,2}$.}  \label{fig:gi4}
\end{center}
\end{figure}
%*************************************************************************************

The pictorial demonstration of the above facts is reported in Fig. \ref{fig:gi3}, where the ground state solutions of the Schr\"odinger equation \eqref{mej10} for both rectangular and smooth potentials from Fig. \ref{fig:gi2} are shown. It is seen that while at $m=4$ the "rectangular" wave function is localized almost between wells (i.e. in this state the wells are coupled substantially), at $m=10$, the wave function shows clear structure with two equal maxima. Latter means that the ground state for $m=10$ signifies weakly coupled wells. This fact is reflected in the corresponding eigenstates, which are both negative and  
$E_0 (m=10) \approx -10.27 << E_0 (m=4) \approx -1.32$, see above. At the same time, for corresponding smooth potentials the wave functions are localized in both wells equally, corresponding to zero eigenvalue at the potential maximum, see Fig. \ref{fig:gi2}. Note that the analytical form of the ground state wave function \eqref{dr1} is identical (indistinguishable in the scale of the plot) to that shown in Fig. \ref{fig:gi3}, which is calculated numerically. This fact could be regarded as a "sanity check" for our numerical calculations. 

The stronger coupling of the wells in smooth (disordered structure) potentials as compared to that in rectangular one (ordered structure) is due to our self-consistency effect. Namely, as we start from the Fokker-Planck diffusion in the confining potential \eqref{mej6}, which "erodes" the initially sharp QW profile, the resulting (necessarily smooth as we use the differentiation in \eqref{mej5} to derive it) profile is determined by the initial random process self-consistently so that the ground electronic states can be obtained exactly (see Eq. \eqref{dr1}) and are distributed equally between the wells. It is clear that this self-consistency is not the case for the approximating rectangular potential, which is not "eroded" by precedent diffusion. This actually means that the disorder is one more factor, promoting the effective coupling of QW's in heterostructures. Our preliminary analysis shows that this is also the case for potentials, consisting of more than two potential wells. 

It is well-known (see, e.g., \cite{qwd}) that while the ground states in each double-well potential in a QW structure can be still separated from each other, the excited states form so-called minibands in real multiple well structure so that latter states play an important role in the physical (especially kinetic like photoconductivity) properties of the semiconducting heterostructures. Two lowest excited states for the cases $m=4$ and $m=10$ are reported in Fig. \ref{fig:gi4}. It is seen that for both analytical (rectangular well profiles) and numerical (smooth well profiles) solutions, the oscillation theorem \cite{land3,bas} holds, which is one more "sanity check" for our calculations. It is also seen that for $n=4$ for $-1<x<1$ the wave functions for rectangular profile give surprisingly good approximation to those for smooth one. This is not the case for $m=10$, which is reflected in eigenenergies $E_1$ and $E_2$ (see above), which are close to each other for $m=4$ only. The fact that the excited state energies $E_1$ and $E_2$ are positive both for $m=4$ and 10 shows that these states realize the strong mixing of those in single wells, i.e. for the excited states the wells in a structure are strongly coupled. For a full understanding of the excited states properties in QW structures, the excitonic effects should be considered. In contrast to bulk semiconductors, excitonic effects are important in heterostructures, including those with QWs especially in their excited states. Full consideration of the excitonic effects in context of our self-consistent disorder model would complicate a lot both initial Fokker-Planck equation \eqref{mej2} and its Schrodinger counterpart \eqref{mej4}. In this case, the problem cannot be reduced to strictly one-dimensional ({\em{ansats}} \eqref{zux1} is no more valid) so that the numerical methods should be heavily used. We postpone the studies of these interesting effects to the future publications.  

\section{Discussion of the experimental manifestations}

The above solution permits calculation of the observable characteristics of the QW structures. One of the important characteristics is optical absorption coefficient, related to the linear optical properties of QW heterostructures. To consider the key features and to demonstrate the possibilities of our method, here we consider the simplest possible model for absorption between the valence and conduction bands in a semiconductor. Namely, we do not consider both excitonic and phonon effects. The latter is because in most common heterostructures like  GaAs/AlGaAs, the electron-phonon interaction is rather weak. To be specific, an electron can raise from the valence band to a state in the conduction band with the same momentum by absorbing a photon. This means that the optical absorption has a form that follows directly from the density of states (DOS) in energy. For bulk semiconductors, the corresponding 3D DOS $g_3(E)$ is given by the formula (see, e.g. \cite{kit,su})

\begin{equation}\label{ds1}
g_3(E)=\frac{\sqrt{2}(m^*)^{\frac 32}}{\pi^2 \hbar^3}\sqrt{E}.
\end{equation}
The result of rigorous calculation of the optical absorption coefficient $\alpha(E)$ within above simple model gives \cite{wilhu}
\begin{equation}\label{ds2}
\alpha(\hbar \omega) \propto g_3(E-E_g) \equiv g_3(\hbar \omega - E_g),
\end{equation}
where $\omega$ is the frequency of an incident photon. Substituting then the discrete eigenvalues $\hbar \omega \equiv E=E_n$ of the Hamiltonian \eqref{mej10} gives the optical absorption coefficient as a series of steps, with one step for each $n$ (number of a localized state in the well). It can be shown, that the corners of the steps join the square root bulk absorption curve \eqref{ds2}. As the dimensional energy of discrete levels in a quantum well scales as the inverse square of its width $E_n \sim 1/L^2$ \cite{land3}, the DOS and optical absorption coefficient would scale as $1/L$. Thus, as the QW width increases, the above steps will become increasingly close to each other until they merge into the continuous absorption curve of the bulk material. 
%*************************************************************************************
\begin{figure}[h]
\begin{center}
\centering
\includegraphics[width=0.99\columnwidth]{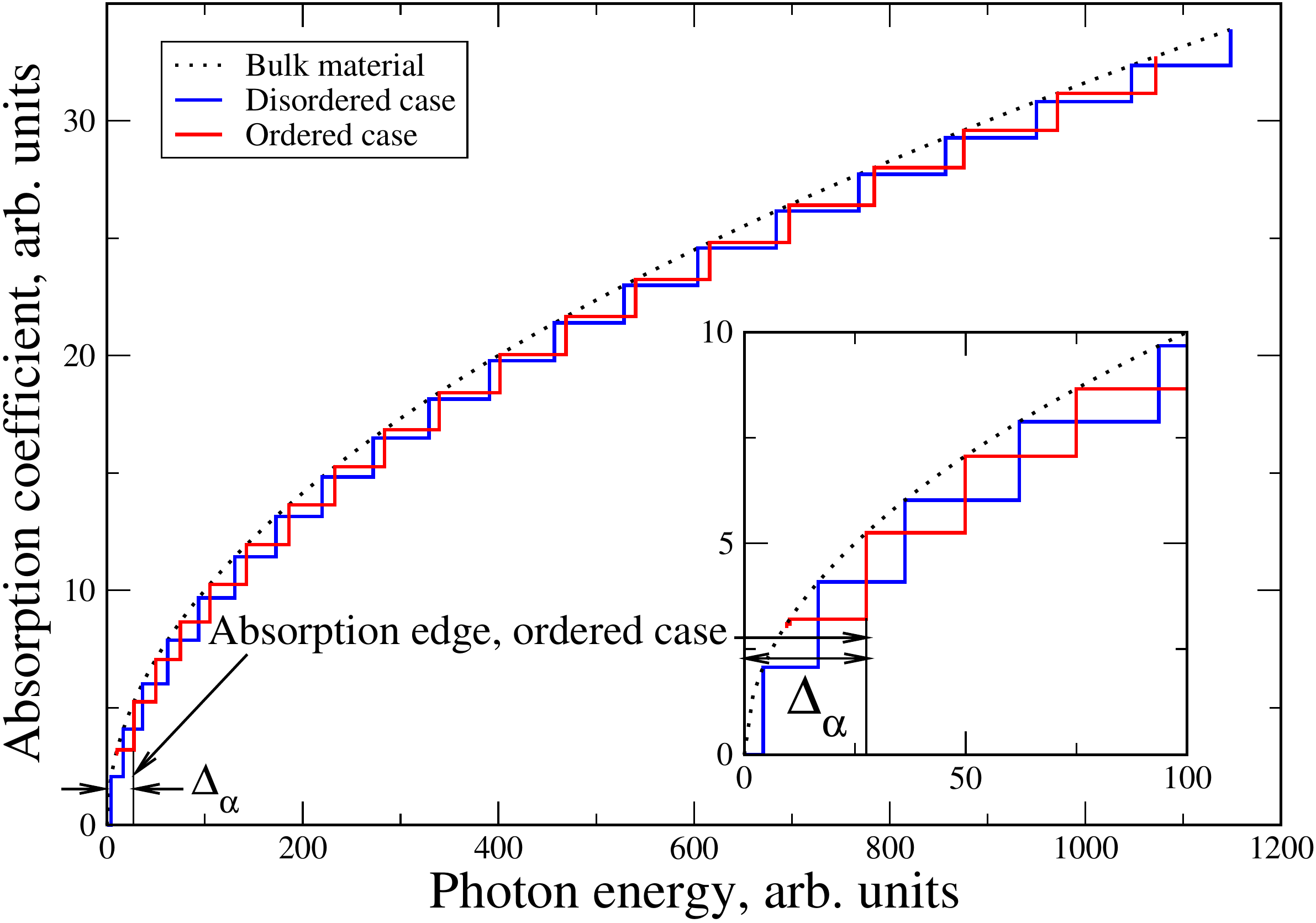}
\caption{Optical absorption coefficient for the case $m=10$ for disordered (smooth QW profile)
and ordered (rectangular profile) cases. The case of bulk material is also reported for reference 
(dotted line). $\Delta_\alpha$ is the difference between the positions of the absorption edges in 
ordered (shown in the panel) and disordered cases. Zero of the horizontal axis corresponds to 
energy gap so that the absorption edge for the disordered case is supposed to be equal to $E_g$.
Inset shows the region near the absorption edge in more details.}  
\label{fig:gi5}
\end{center}
\end{figure}
%*************************************************************************************
The optical absorption coefficient $\alpha$ for the potentials with $m=10$ and $\kappa_m=m^2$ is reported in Fig. \ref{fig:gi5}. 
The behavior of $\alpha$ is qualitatively the same for the potentials with other $m$ and $\kappa_m$. As we discussed above, the optical absorption coefficient is proportional to DOS and comprises the sequence of steps, signifying the levels in the QW or so-called subbands. The latter name stems from the fact that in a QW layered structure the charge carriers are still free to move in the directions parallel to the layers. This means that we do not really have discrete energy states in QW, rather we have  "subbands" that start at the energies calculated for the above discrete states. 

To plot the optical absorption coefficient \eqref{ds2} in Fig. \ref{fig:gi5}, we took 21 first eigenvalues for both smooth and rectangular wells. It is seen that while for "diffusion-spoiled", i.e. smooth QW profile, the absorption starts exactly from $E_g$, in the ordered case, the absorption edge is shifted to the value $\Delta_\alpha$, which for the above parameters is around 27 in dimensionless units. In a typical GaAs QW structure we have the electron effective mass $m^*\approx 0.067 m_0$ ($m_0$ is a free electron mass) and well width $L=10$ nm, see, e.g. \cite{pp}. For these data we obtain that $\Delta_\alpha \approx 50 $ meV, which lies in the infrared region of frequencies.  On the other hand, the absorption coefficient has been measured in Ref. \cite{njp} for the Ge multiple QW structure, see Ref. \cite{njp} for the details of the structure fabrication. The linear absorption spectrum (which is the case for our theory) is reported in Fig. 2 of the Ref. \cite{njp}. To compare our results with those of \cite{njp}, we observe that our calculations correspond to zero temperature. This means that it is reasonable to compare them with the experimental curve for $T=10$ K in Fig.2 of Ref. \cite{njp}. Taking the effective mass value for Ge $m^*\approx 0.12 m_0$ \cite{pp} and QW width $L \approx 10$ nm (see the right panel of Fig.2 of Ref. \cite{njp}), we obtain that the position of the first peak around 0.96 eV corresponds to approximately 200 in the dimensional units of our Fig. \ref{fig:gi5}. Also, for the above QW width, our scale of the dimensionless absorption coefficient $\alpha$ corresponds to that on the vertical axis of the experimental figure. This gives that for the first peak of absorption, the experimental value is around 12, while the theoretical one is approximately 14, which gives the error around 16\%. Such error (for the uncertainties in the parameters like effective mass, QW width, etc) is not bad for the theory, which does not account for many-body effects like in DFT approximations.

In relation to the above, we note that small variation of parameters $L$ and $m^*$ can easily drive the absorption edge shift to the visible and ultraviolet ranges of wavelengths. This is because the energy level spacing in QW grows for wells narrowing (i.e. $L$ decreasing) and an effective mass $m^*$ reduction. This means that QW structures are quite sensitive to the QW profiles disordering due to diffusion. This fact should be taken into account while designing the optoelectronic devices, based on these structures. This is because at elevated temperature such diffusion in certain time span can detrimentally influence on such device functionality.    

\section{Conclusions and contexts.}

Here we demonstrated the possibilities of the suggested "self-consistent disorder" method for the calculation of the observable properties of QW semiconductor structures. Although we calculated the optical absorption coefficient only, the method is capable to calculate virtually any characteristic of the QW structure. Our approach is unique as it permits us to consider disorder effects self-consistently, which is impossible in DFT methods. As we have seen above, the numerical calculations in our method require much less computational effort as compared to that in DFT approaches. At the same time, our theory can describe the experimentally observed quantities like optical absorption coefficient (see above) with an accuracy of around 16\%, which is rather good for the effective single-particle approximation. The method can be easily generalized to the account for external electric and magnetic fields (corresponding terms can be added to the Hamiltonian \eqref{mej1b} like it had been done in DFT simulations for InN/GaN multi-well structures \cite{pol}) so that we shall be able to calculate the electro- and magneto-optical properties of QW structures concerning their amorphization. This shows the ubiquity of the method, which does not rely upon the empirical approximations of the alloy's lattice constants, bandgaps and other parameters \cite{dftpccp}.

The second obvious generalization of the method suggested is the account for the excitonic effects. The direct consideration of these effects reduces to adding up a Coulomb (also screened one, see, e.g. \cite{cud}) interaction to the potential $V({\bf r})$ \eqref{mej5}. This, however, will change the initial "diffusion" potential $U({\bf r})$ in a self-consistent way. Such a problem can barely be solved analytically. Rather, the direct numerical solution of 3D Fokker-Planck and Shcr\"odinger equations should be necessary in this case. Also, the optical absorption coefficient will not only involve the direct excitation of an electron from the valence to the conduction band but also the creation of electron-hole pairs, i.e. excitons. This makes the calculations even more involved as the altering of the entire band structure by the excitonic effects should be considered.  Our study shows that the inclusion of excitonic effects especially augmented by Rashba spin-orbit interaction (SOI) \cite{rsh} can explain some salient experimental features in QW structures like additional sets of peaks in the optical absorption spectra. The role of SOI in the QW structures with fluctuating dopant concentration in the QWs sides had been considered in Ref. \cite{sher03}. It had been shown that such randomness causes finite spin relaxation rate (contrary to the "ordered" case where the rate is infinite), which can be visible in the above spectra. On the other hand, for ordered case, the fully fledged consideration of the QW structures with respect to SOI had been done in the Ref. \cite{ufla} within the DFT theory. As usually for DFT approaches, the many-body effects has been taken into account through the exchange-correlation term \cite{ufla}. This already made the consideration very cumbersome and computationally demanding. This is because the solution even of our problem with respect to SOI is much more complicated as the electron wave function becomes spinor in this case \cite{grim}. Nevertheless, in our case the problem might be doable in momentum space similar to Ref. \cite{my18}. This shows the relative simplicity (for sure at the level of numerical computations) of our self-consistent formalism, which can give (see above) the satisfactory description of experimental results without cumbersome explicit consideration of many-body effects.

Note that our method of disorder handling is naturally (as it is usual in the formalisms dealing with QWs) continuous so that it does not take into account the initial lattice discreteness. However, it is well known that the disorder is indeed a lack of regularity. This means that a disordered substance (especially for strong disorder leading to the amorphization) has its atoms not arranged periodically. In other words, the lattice constant value will fluctuate. The amplitude of such fluctuations in our formalism can be conveniently described by the Langevin stochastic equation. In this case, the potential, related to the lattice constant fluctuations, will contribute to that in our "random diffusion" mechanism, see Eq. \eqref{mej3}. This implies that the corresponding Fokker-Planck and Schr\"odinger equations will acquire additional potential. Our analysis shows that as our "random diffusion" potential is a fast - growing function, the above lattice constant fluctuation contribution will not change our results quantitatively. In this context, our method can be generalized to non-Gaussian random processes by introduction the fractional derivatives \cite{kilbas} into the Fokker-Planck \cite{kilbas,sok1,obz} and Schr\"odinger \cite{lask,laskin} equations. Latter generalization leads to so-called fractional quantum mechanics \cite{laskin}, which can be used to consider not only quantum well properties \cite{yama} but also the excitonic effects in them \cite{my2019,my2020}. Latter effects become especially interesting in the presence of Coulomb interaction screening both in ordered \cite{cud} and disordered case \cite{scirep}. This is important since the screening defines many optical and transport properties of layered semiconductor devices so that knowledge of the disorder's influence on its behavior is fundamental for practical applications. For instance, the recently synthesized MoS$_2$ monolayer could be used as a single-layer transistor \cite{nnat} thanks to the confining properties of its low-dimensionality and Coulomb interaction screening.

Perhaps the main conclusion of this work is that the disorder, especially in considered here self-consistent form, leads to abrupt shifting of the initial (i.e. "sharp") energy levels thus altering (sometimes substantially) the physical characteristics (like light absorption edge) of the layered semiconductor structures. Further experimental and theoretical work on the joint influence of excitonic effects (along with Coulomb interaction screening) and SOI is required for a complete understanding of the charge carriers properties in disordered low-dimensional semiconductor structures. 

 \begin{acknowledgments}
We are grateful to E.Ya. Sherman for discussions. This work was supported by the National Science Center in Poland as a research project No.~DEC-2017/27/B/ST3/02881.
\end{acknowledgments}

\end{document}